\documentclass[aip,rsi, reprint]{revtex4-1}

\usepackage{graphicx}
\usepackage{dcolumn}
\usepackage{bm}

\usepackage[english]{babel}

\usepackage{url}

\usepackage{epstopdf}

\begin{document}
\title{Investigations of dynamic light scattering properties in fluorescent solution: A comprehensive study using plasmonic enhanced Flow-Cytometry}

\author{Martin Michelswirth}
\affiliation{Attosecond Research and Science Group, CFEL science, Hamburg University, http://ars.cfel.de} 
\author{Julia Hengster}
\affiliation{Attosecond Research and Science Group, CFEL science, Hamburg University, http://ars.cfel.de} 
\author{Tobias Milde} 
\affiliation{Department of Biomaterials, Institute for Bioprocessing and Analytical Measurement Techniques, Heilbad Heiligenstadt, http://www.iba-heiligenstadt.de/en/fachbereiche/biowerkstoffe/} 
\author{Thomas Gaumnitz} 
\affiliation{Attosecond Research and Science Group, CFEL science, Hamburg University, http://ars.cfel.de} 
\author{Natalie Wagner}
\affiliation{Institute for Technical and Macromolecular Chemistry, Hamburg University, http://www.chemie.uni-hamburg.de/tmc/theato/index.html}
\author{Patrick Theato}
\affiliation{Institute for Technical and Macromolecular Chemistry, Hamburg University, http://www.chemie.uni-hamburg.de/tmc/theato/index.html}
\author{Klaus Liefeith}
\affiliation{Department of Biomaterials, Institute for Bioprocessing and Analytical Measurement Techniques, Heilbad Heiligenstadt, http://www.iba-heiligenstadt.de/en/fachbereiche/biowerkstoffe/} 
\author{Thorsten Uphues}
\affiliation{Attosecond Research and Science Group, CFEL science, Hamburg University, http://ars.cfel.de} 

\begin{abstract} 
In this article we present a new diagnostic approach utilizing flow-cytometry to study compounds of nanoparticle samples in solution by analysis of their scattering patterns retrieved from the cytometric measurements. As a specific enhancement of this technique we study as well the scattering pattern of nanoparticles in a fluorescent solution (529 nm). A significant enhancement of the cytometry measurements is observed supporting an improved separation of particle formations that are clearly resolved in the cytograms. The samples in this experiment are prepared from 80 nm citrate-capped gold nanoparticles (AuNP). They are stabilized providing 2-(dimethylamino)ethanol (DMAE) in the aqueous solution. A laser diode with a wavelength of 488 nm is used as fundamental illumination for the flow-cytometry measurements (FCM). Dynamic Light Scattering (DLS) measurements are performed separately and demonstrate a very good agreement with the Flow-Cytometry  measurements both of which allow to give an effective size calibration. For further analysis light transport simulations are presented. They provide information on the key-process to form the correlation of the fluorescent solutions FCM to the studied particles of interest. From this we extract the volume nature of the scattering process ensuring the correlation.
\end{abstract}

\maketitle

\section{Introduction}

Current microbial studies drive innovative developments leading to various clinical and microbiological diagnostic applications like flow cytometry \cite{Alvarez-Barrientos2000,Pedraza-reyes2014,Zuleta2014} which is of particular interest for the new approach followed in this article. Since the development of flow cytometry with the first apparatus  in 1965\cite{Kamentsky29101965} these devices were constantly developed to feature diagnostic applications with increasing performance.

Cell diagnostics analysing forward and sideward optical scattering properties are established by flow-cytometry techniques without any staining of the sample solutions. This gives rise to developments of functional staining reactants and fluorescent detection techniques to address scientific questions with respect to cell-activity and -vitality. In particular investigating the cell esterase activity is driving research from the beginning of these apparative and functional improvements \cite{Norden1995} in cytometry techniques and is still a major topic of recent scientific and clinical interest \cite{Hyka2013,IJC:IJC29155}.
According to the new insights arising from nano applications in medicine this technique is of increasing interest with respect to the optical properties of tailored nanoparticles in medical applications. This faces new developments in Flow-Cytometry (FCM) for different types of applications. As a brief overview fast automated cell sorting \cite{Jaye2012}, in vivo \cite{IJC:IJC29155,Goode2009}, imaging \cite{Barteneva2012} and real-time flowcytometry \cite{Rebelo2013, Zuleta2014} techniques are already utilized to perform specific tumor cell medical studies \cite{Greve2012,MacLaughlin2015,Manju2012,Xu2013,Galanzha2008,Hsu2009}.
Besides other topics in the application of Nanoparticles (NP) in medical applications, Nanoparticle assisted diagnostic and therapy approaches provide a huge variety of FCM applications for cancer- and stem cell research. Plasmonic enhancement is already a key technique to improve material contrast \cite{Galanzha2008, Jensen2011, Lim2011} in diagnostic and therapeutic fields and could be used in FCM as well. Localized excitation of gold (Au) NPs enables transport of anticancer drugs \cite{Manju2012,Xu2013,Galanzha2008,Hsu2009}, where the cellular specific reception is precisely specified by the parameter space extracted from flow-cytometry \cite{MacLaughlin2015,Manju2012,Xu2013} measurements. To our best knowledge current research and the related therapy approaches of this kind are still necessarily combined with a fluorescence labelling of the AuNPs. 
Especially multiple cell targeting \cite{Brown2010} utilizing AuNPs is  limited by this cell specific labelling. Thus FCM applications are usually confined on the conventional cell stained vitality \cite{Greve2012} or the cell specific reception quantifications \cite{MacLaughlin2015,Manju2012,Xu2013}. 
In this article we present a novel FCM scheme, where no fluorescent labelling of the AuNPs is needed at all. We demonstrate that our specific amendments do not change the standard process of FCM but can significantly enhance the amount of information gathered from these measurements. 

Therefore, our approach focuses on the utilization of localized ambient illumination of the nanoparticle in solution. This diffuse illumination originates from fluorescent emission of the solvent medium on a molecular scale which we call a volume illumination effect further on. As a consequence the response of the investigated particles is a localized light scattering information inside the volume. This response provides additional and so far untouched information gathered by Flow-Cytometry especially in the sideward detection direction. We demonstrate that plasmonic enhancement improves the contrast separating the particle scattering from solvent ambient emissions even under presence of a massive unspecific scattering background. The nature of volume light transport associates the particle scattering response to the localized ambient fluorescent excitation apart from the diffuse emission.

We can clearly resolve an enhanced particle response in the sideward scattering signature, which can be detected with fluorescent flow-cytometry \textit{(FL/FSC)}. 
In the following we discuss an improved specificity for the detection of the particle of interest which can lead to new applications investigating unspecific and unknown sample solutions in terms of plasmonic enhanced solutions without the necessity of fluorescent labelling of the particles themselves. We emphasize that for all measurements presented in the following not fluorescent labelling of a particle was existing or initiated. Only the ambient illumination in the solution is supported by a fluorescent excitation at\textit{ }529 nm caused by the DMAE assisted reactions with the solvent medium.
For a more extended study of this new technique we force agglomeration of the particles changing the solvent environment by addition of NaCl \textit{(aqueous, 0.9 \%)} into the stabilized solutions. The flow-cytometric studied samples are characterized by Dynamic Light Scattering (DLS) in more detail. A characterization of these localized particle formations is only discussed briefly based on the \textit{FL/FSC} Flow-Cytometry observations.
In addition to this new approach in FCM, FTIR studies are performed in solutions of the DMAE stabilized citrate-capped AuNPs. For the FTIR measurements a fluorescein based labelling of the compounds is prepared to determine whether such labellings are inhibited by the provided amine surplus \textit{(1.7 mM, DMAE)}. These measurements show that a stimulated labelling is prevented, if the nanoparticle solutions are not forced to agglomerate.

\section{Experiment}

\subsection{Material and Methods}

\subsubsection{Preparation}

Citrate capped gold nanoparticles in aqueous solution are stabilized providing 2-(dimethylamino)ethanol (DMAE) at 50$^\circ$C (oil-bath) temperature under Ar atmosphere. Solutions were stirred at this temperature for 90 minutes.\\
The chemical reduction driven due to the protonation of the tertiary amine Nitrogen functionality is performed at a low concentration level of \textit{0.17 m}\textit{M} of DMAE and a reaction proceeding in the presence of \textit{1.7 mM} of DMAE is performed to obtain a mono-disperse sample solution. Agglomerated samples are prepared by a provoked charging of the amine stabilized particle solution by adding aqueous \textit{NaCl} \textit{(80 mM)}. A negative control is prepared repeating the procedure by use of \textit{1.7 mM} of DMAE agent in absence of the gold particles. For the negative control sample the nanoparticle solution is replaced by ultra-pure water \textit{(Millipore)}. 

All glassware is cleaned with ultra-dry (alkali metals processed) ethanol and dried at \textit{120 }$^\circ$C for \textit{15 minutes}. All reactants are added under Schlenk conditions. The gold nanoparticles \textit{(CANdots Series G, 0.1 mg/ml)} are used in aqueous solutions as purchased (CAN GmbH, Hamburg, Germany). These solutions were degassed several times before adding the DMAE agent. The solvent amount of the purchased particles is retained in spite of a dilution in the case of the mono-disperse reference sample. The \textit{NaCl}  is added as \textit{0.9\%} aqueous solution.

\subsubsection{Flow-Cytometry}

Measurements are performed with a \textit{CyFlow Space} (Partec, PMU) Flow-Cytometer. The fundamental excitation and illumination wavelength is \textit{488(}$\pm$\textit{2) nm} (50mW, laser diode). The sideward detection is realized by a \textit{x50} objective lens \textit{(NA 0.82, WD 0.4 mm)} for collimation. The fluorescent response from scattering in the sample solutions are analysed in sideward detection \textit{(SSC)} passing interferometric optical filters (\textit{536/40 nm} and a \textit{488/6 nm}). The detector gains are adjusted to be \textit{x2.4}- (at fluorescence, \textit{FL}) and \textit{x0.8} (sideward scattering, \textit{SSC}) with respect to the forward detection\textit{ (FSC)}. The \textit{FL} and \textit{SSC} flow-cytometry measurements are performed simultaneously.

\subsection{Procedure}

\subsubsection{Colocalization Analysis}

The cytograms are recorded as histograms with \textit{12-bit} (4096 x 4096 channels) resolution. They are analysed performing a 2-dimensional correlation analysis. Density plots are generated with a scaling to \textit{400 x 400 bins} (no interpolation). The colocalization cytogram plots are retrieved comparing the \textit{SSC/FSC} and the \textit{FL/FSC} bi-variable density data. This analysis is performed using a \textit{'}\textit{RG2B Colocalization}\textit{'}\textit{ }code.\cite{RG2B,Shapiro2005,Hida2012} The \textit{FL/FSC} to \textit{SSC/FSC} cytograms colocalizations are scaled to exploit the 8-bit pseudo-color scale, respectively. The event axes in all cytogram plots are presented as measured and are scaled in units of "thousand" channels.

\subsubsection{Volume scattering calculations}

 Light transport simulations are performed based on Monte-Carlo (MC) calculations. The volume scattering functions are shaped by \textit{Henyey-Greenstein (HG) }\cite{HG:1941,Kienle2004,Gao_Huang_Yang_Kattawar_2013} analytical phases. The defuse fluorescence is supposed to be caused by isotropic scattering events \textit{(HG, g=0),} where the scattering in the sample cell volume is introduced by using a \textit{g=0.64} (forward/backward) anisotropy parameter. As the coefficient \textit{}of attenuation by scattering a value of \textit{$\mu$}\textit{$_{s}$}\textit{=139.861 1/cm} is used. This value is assigned to the scattering events and is retrieved from MIE-calculations, where an effective concentration of three primary gold nanoparticles \textit{(80 nm) per cubic micro}\textit{n} is assumed. The HG anisotropy factor \textit{g} is obtained by stimulated agglomeration to effective sizes of \textit{1830 nm} and used to estimate the optical scattering parameters in the pre-calculations. In all calculations aqueous solutions are assumed. The optical refractive indices in MIE-calculation are obtained from \cite{Johnson1972}, where the real one of the medium (supposed to be pure water here) is used from \cite{Hale1973}.\\
 
\begin{figure}[t!]
\includegraphics*[width=\columnwidth]{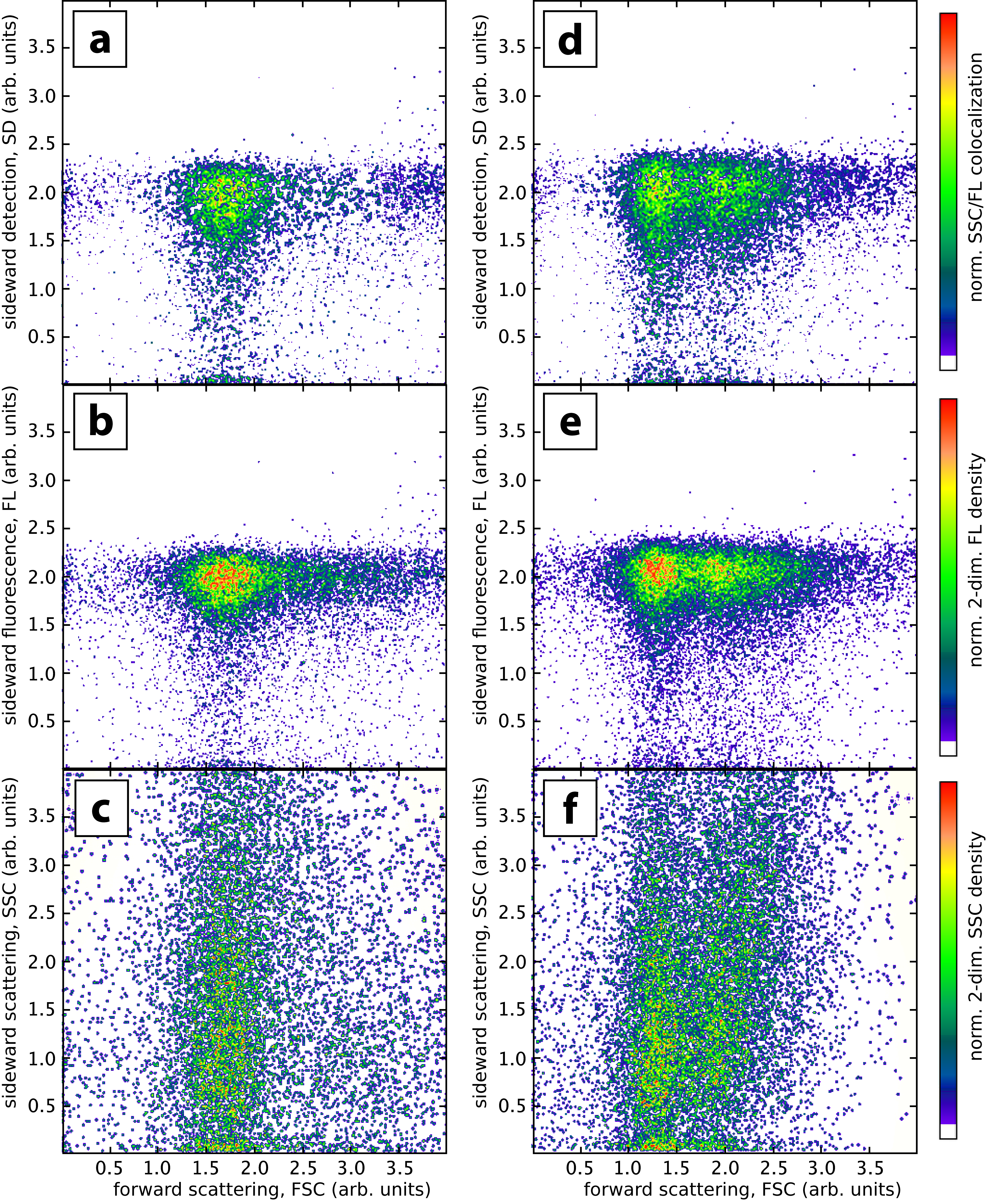} 
\caption{\footnotesize{Cytograms of DMAE reduced citrate-capped AuNP solutions: Reactions in presence of 1.7 mM of DMAE (a, b, c) in a direct comparison to a poly-disperse formation due to charged capping provoked agglomeration, when adding 0.9\% of NaCl (d, e, f). Colocalization plot (a, d), fluorescence cytograms (b, e) and of the pure scattering measurement (c, f), respectively.
}}
  \label{fig:abb1}
\end{figure}
 
The sample volume cell used in the volume light transport simulations is defined by a 3-dimensional mesh of \textit{444x444x444}. With these simulations both cases model scattering events, thus the volume fluorescence (g=0) and fundamental scattering (g=0.64) can be compared. The dimensions of the model cell is given by a cube of \textit{7.4 $\mu$m} length. The walls are set to support no further boundary condition. A radial symmetry is supposed averaging over transversal directions. To cover the cell volume most efficiently the transport simulation has been repeated six times. The initial trajectories at the illuminating spot are randomly distributed. About \textit{13-24 thousand }photons per iteration are simulated in the \textit{g=0.64} (scattering) and \textit{18-32 thousand} photons per iteration are simulated in the \textit{g=0} (isotropic, fluorescence). A detailed view of the modeled cell spanning a region of \textit{1.3 $\mu$m x 1.3 $\mu$m }and \textit{0.65 $\mu$m} (in propagation direction) is observed to cover all trajectories information, which does not vanish after averaging over six iterations with \textit{(8-bit)} resolution.\\
The illumination is limited to a size of \textit{1 $\mu$m} (Gaussian and placed \textit{0.03 $\mu$m} behind the entrance surface. The MC calculations are performed by using a light transport simulation algorithm \cite{mcxyz,Ramella-Roman2005a,doi:10.1117/1.2967535,Kienle2004}, which basically yields fluence normalized to the total throughput. The forward propagated relative flux rates obtained are converted to logarithmic scaled measurements of attenuation giving an impression of their spatial distributions as they differ for the volume fluorescence and the volume scattering events.\\
The retrieved maps of attenuation by fluorescence or by scattering caused from the investigated localized particle formation are analysed with respect to the sideward direction. This direction corresponds to the direction of the flow-cytometry measurements.\\
The simulations are aimed to get an idea, whether scattering caused by quasi localized particles in solution affects the cytometric detectable response of the sample solution in a sufficient way, to provide significant dissimilarities to a pure and diffuse volume fluorescence.

\section{Results and Discussion}

\subsection{Flow-Cytometry }

After the amine reduction by chemical preparation the citrate capped AuNP solutions (primary particle size about \textit{80 nm}) are investigated in a Flow Cytometry setup. The samples are studied in fundamental \textit{(488 nm)} scattering cytogram \textit{(SSC/FSC)} acquisition.\\
The results are compared to the fluorescence detection measurements (FL/FSC). A FL (529 nm) specific detection is ensured by the utilized optical filter. A spectral interference of fundamental scattering and fluorescence is strictly avoided with respect to the spectral characteristics of the interference filters \textit{(488/6 and 536/40)} used in sideward detection.\\
Fig. \ref{fig:abb1} shows an analysis of correlations (Fig. \ref{fig:abb1}a) for the \textit{1.7 mM} stabilized AuNP solution , the \textit{FL/FSC} (Fig. \ref{fig:abb1}b) and the \textit{SSC/FSC} cytograms (Fig. \ref{fig:abb1}c)of an agglomerated sample solution (see Fig. \ref{fig:abb1}d, e, f). The false-colors of the correlation cytogram plots correspond to the ratios of colocalization of the analysed FL/FSC and SSC/FSC measurements, respectively. Both of these measurements are analysed with respect to colocalizations in the equally scaled fluorescence and scattering cytogram plots. The scattering and fluorescence cytograms are presented as density plots. The event axes are scaled by 1/1000 with respect to the measured channels (12-bit resolution).

For the \textit{1.7 mM} DMAE AuNP sample solution a single feature is observed in the \textit{SSC/FSC} cytogram (Fig. \ref{fig:abb1}c). It is slightly tilted with an angle of \textit{3-4 deg}. A significantly enhanced separation from the cytogram background is obtained from the corresponding \textit{FL/FSC} (b) measurement in contrast to the conventional fundamental scattering \textit{SSC/FSC} (c). An unspecific scattering background is found to interfere with the \textit{SSC}  measurement.\\
The \textit{1.7 mM} DMAE AuNP sample solution is analyzed to be highly monodisperse with respect to the \textit{FL} measurement.\\
The feature, which is associated to the corresponding formation of particles in solution, is localized at about \textit{1.74} (\textit{FSC}-axis projection). This is retrieved to be colocalized in both of the measurements, the \textit{SSC} and the \textit{FL} detection (compare a)). A detailed analysis of dispersities of the compared sample solutions is presented later. In the right side panel of Fig. \ref{fig:abb1} d, e, f the same measurements but with agglomerated AuNP solution are shown and compared to the monodisperse AuNP sample.

As for the monodisperse sample solution the \textit{FL} detection cytogram measurements (e) show a clearly improved separation from the cytogram background for the agglomerated AuNP sample. The agglomeration of the AuNPs is found at a \textit{FSC 1.9-2.2} localized feature in our measurements. A weaker \textit{1.7 FSC} remaining feature is re-established in the analysis of both the \textit{FL} and \textit{SSC} detection, when their colocalizations in the cytogram plots are analyzed. The colocalization plot is shown in Fig.1d. It is presented in comparison to the corresponding plot of the measurements of the stabilized monodisperse sample solution (see the counterpart left panel a)). In addition to the \textit{FSC 1.9-2.2} feature (Fig. \ref{fig:abb1}d) the agglomerated particles in the \textit{NaCl} enriched solution appears in a increased tilt of the SSC cytogram measurement (Fig. \ref{fig:abb1}f).  It is found to be \textit{7-8 deg} in contrast to the \textit{3-4 deg} for the stabilized sample.

This already gives a first information on the corresponding agglomerated sizes. The increase of the tilt towards the diagonal of the feature at \textit{FSC 1.9-2.2} in the \textit{SSC/FSC} plot can be directly associated to the formation of bigger agglomerates. The effective sizes of the agglomerates are still in a regime where the scattering events are not dominated by elastic optical scattering since elastic scattering driven cytogram measurement would show a distinct alignment of the respective size associated features along the diagonal of the \textit{SSC/FSC} plot.\\

\begin{figure}
\includegraphics*[width=\columnwidth]{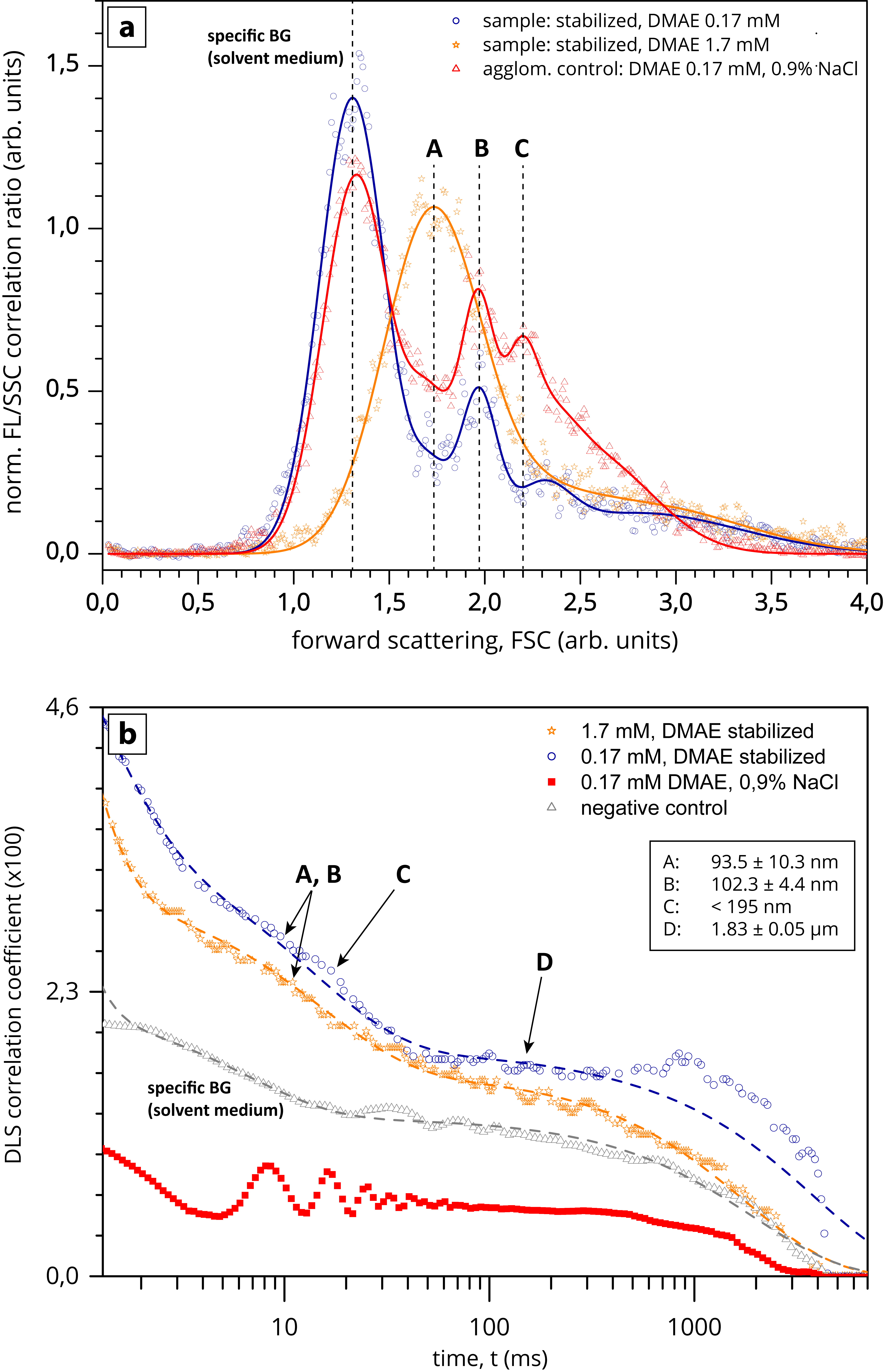} 
  \caption{\footnotesize{Fraction analysis (a) of DMAE reduced citrate-capped AuNP solutions: 0.17 mM DMAE stabilized (blue), 1.7 mM DMAE stabilized (orange) and the 0.9\% NaCl added agglomeration control one (red). Calibration DLS measurements (b): 0.17 mM DMAE stabilized (blue), 1.7 mM DMAE stabilized (orange), negative control (no AuNPs) (grey)  and the corresponding fits (dashed).
}}
  \label{fig:abb2}
\end{figure}

In Fig. \ref{fig:abb2}a a detailed analysis of the cytogram measurements is given. Profiles obtained by averaging along the colocalizations axis of the measurements (Fig1a,d) are presented as they are obtained from the \textit{FL} to \textit{SSC} colocalization analysis plots. 

An additional measurement for a very low DMAE \textit{(0.17 mM concentration AuNP sample solution)} is given for comparative reasons in this diagram. All profiles retrieved are analyzed performing fits based on Gaussian profiles. The profiles of all the three sample types are shown with the corresponding fit.\\

From the fit of the profiles associated to the cytometry measurements of the \textit{1.7 mM} DMAE stabilized sample we can see that this solution is verified to show a high degree of mono-dispersity (see orange curve). No further feature apart from the one localized at \textit{1.726}\textit{$\pm$}\textit{0.002 FSC} ("A") is observed. The width of this feature is found to be \textit{0.301}\textit{$\pm$}\textit{0.003 (FSC)}. For the forced agglomeration (blue curve) the dispersity turns out to show a more complex situation. The associated colocalization cytogram profile is matched by a 4-components fit. In addition to the feature "A", which is still present, additional features localized at \textit{1.966}\textit{$\pm$}\textit{0.006 FSC} ("B") and \textit{2.184}\textit{$\pm$}\textit{0.008 FSC} ("C") and a further component at \textit{1.328}\textit{$\pm$}\textit{0.004 FSC }show up.\\
It is pointed out here, that there is no cytogram component observed in the region of the \textit{1.328} \textit{FSC}, if the solution is stabilized by the \textit{1.7 mM} DMAE. This component is found to turn up in the \textit{0.17 mM} DMAE AuNP sample (dark-green curve) as well as the component "B".

A further feature is localized at \textit{2.296}\textit{$\pm$}\textit{0.018 FSC}. It is observed to be slightly shifted with respect to "C". An additional component at \textit{2.805}\textit{$\pm$}\textit{0.038}  is associated to the feature of the highest FSC projection value for the very low DMAE concentration solution.

Additional MIE-calculations are performed to support the evidence of a plasmonic resonance enhancement as it occurs in our cytometry configuration for all measurements at \textit{488 nm} excitation. For effective particle sizes up-to \textit{225-300 nm} a plasmonic enhancement of all cytometry measurements is rectified if the effective particle formations have sizes smaller than this (see ESI).

To verify the \textit{1.328 FSC} feature measured in the agglomerated sample a particle size characterization of the particle solutions by Dynamic Light Scattering (DLS) is performed. The measurements are shown in Fig. \ref{fig:abb2}b. The \textit{1.7 mM} DMAE stabilized solution (orange) and the low concentration \textit{0.17 mM} modified solution (dark green) as well as a negative control containing no particles (light gray) are measured. The recorded correlation time-traces are analyzed by a multi-component fitting procedure. Relaxation rates are extracted from this analysis of the DLS measurements. The hydrodynamic sizes in diameter are calculated from the translation diffusion coefficients. Fitting for the agglomerated sample (blue curve) was not possible due to the low correlation time-trace quality of the measurement.

The DLS measurements were recorded under \textit{173 deg} and therefore a backscattering configuration utilizing an illuminating laser wavelength of \textit{632 nm} is used. This configuration serves for a sufficient scattering signal for our aqueous sample solution, if the containing AuNP formation shows effective sizes smaller than \textit{300 nm}. (see blue curve obtained from the MIE-calculations, ESI2).\\
Nevertheless, the hydrodynamic sizes retrieved from the DLS measurements of our sample solutions are in good agreement with the results from the flow-cytometry colocalization plots.\\
Thus, the feature "A" and therefore the monodisperse sample solution is nicely matched to an effective hydrodynamic size of \textit{93.5}\textit{$\pm$}\textit{10.3 nm }from the DLS calibration measurement. \\

The components associated to the mean sizes of the particle formations in our sample solutions retrieved from DLS are assigned to the observed cytogram features. Different sensitivities with respect to the sizes of contributing particle formation in the cytometry and the DLS measurements are deduced from the performed MIE calculations (see the red and blue curves in ESI2).
The sub-dimer effective size associated to the \textit{1.7 mM} solution component "A" represents an agglomeration of less than two NPs and is even valid for multicomponent fitting analysis of the corresponding DLS measurement.\\
Features ("A","B","C") as they are revealed from the colocalization analyzed cytometry measurements (\textit{1.726 FSC, 1.966 FCS and 2.184 FCS} projection) of our AuNP sample solutions are matched to the DLS compounds and their effective hydrodynamic sizes (\textit{94 nm, 102 nm and 180 nm}), respectively. In addition to the compound "A", which is reproduced with lower intensity in the agglomerated AuNPs solution cytometry measurements, two further compounds are associated to the common cytogram features from the agglomerated and the \textit{0.17 mM} modified solution.\\

These compounds "B" and "C" are successfully matched to the correlation measured in the DLS traces for the \textit{0.17 mM} DMAE modified sample. Compound "C" is strongly influenced by the scattering background of the DLS measurement of this sample caused by the medium. This is related to the complexity of the impure AuNP solutions and the reduced scattering efficiency of bigger agglomerates.\\
Following the corresponding time trace from the DLS measurements at values above roughly \textit{100 ms} for the \textit{0.17 mM} DMAE modified solution a single component fit model is used. It is associated to any bigger agglomerated compounds with effective sizes \textit{>300 nm}. These are subsumed with a component "D" associated with a hydrodynamic size of \textit{1.83}\textit{$\pm$}\textit{0.05 $\mu$m}.\\
A common compound is revealed by the DLS measurements of the negative control sample solution (no NPs) and the \textit{0.17 mM} low DMAE concentration modified AuNP solution.\\
Its appearance in the negative control sample without AuNPs associates this compound to the solvent medium impurities. This is seen to correspond to the cytometry colocalizations observed at the lowest FSC-axis value projected localized feature at \textit{1.31 FSC}. This feature is detected for the low concentration DMAE (\textit{0.17 mM}) sample as well as for the agglomerated control sample. It turns out to vanish completely for the monodisperse solution cytometry measurement (\textit{1.7 mM} DMAE). This flow-cytometry result corresponds to a compound size near to the plasmonic resonance at the exciting wavelength of the FLM device ("A" is associated to a size of \textit{93.5 nm}) and of the formed homogeneity of this solution.\\

%

\begin{figure*}[t!]
\begin{center}
\includegraphics*[width=0.85\textwidth]{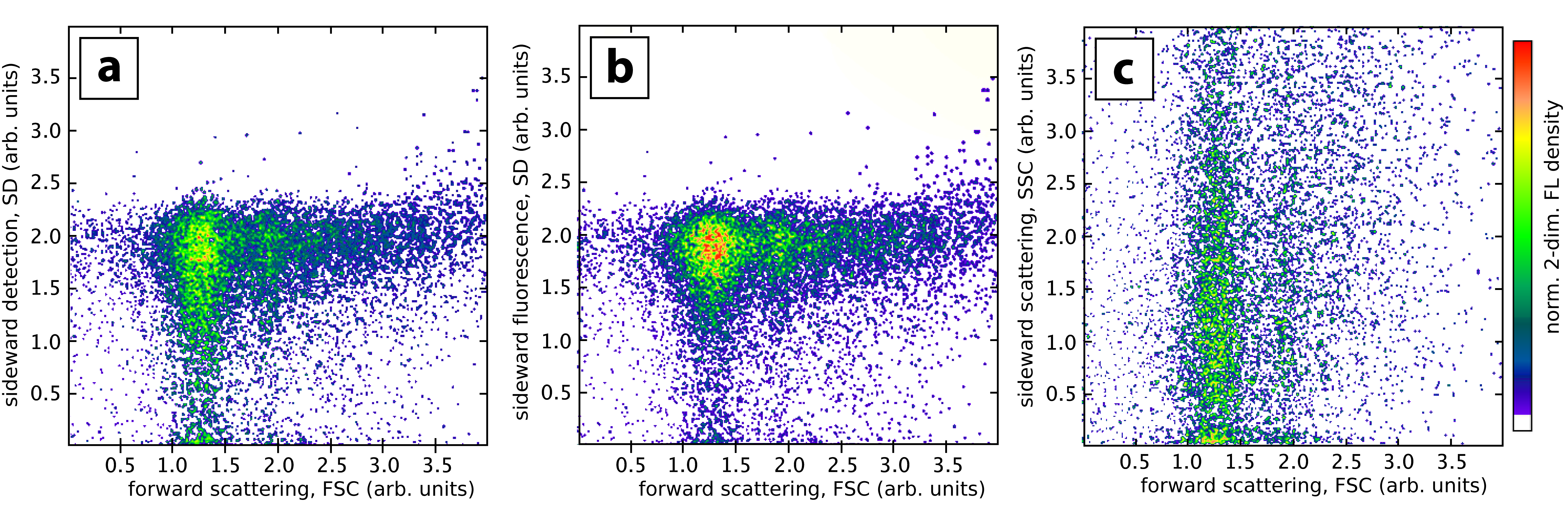} 
\end{center}
  \caption{\footnotesize{Cytograms of DMAE reduced citrate-capped AuNP solutions: Reactions in the presents of a low  DMAE concentration of 0.17 mM . Colocalization plot (a), fluorescence cytograms (b) and of the pure scattering measurement (c).
}}
  \label{fig:abb3}
\end{figure*}

Finally it is demonstrated that the AuNP \textit{1.7 mM} DMAE stabilized solution shows a homogenization in the AuNP particle formation size. The associated effective size is found to be about \textit{93.5}\textit{$\pm$}\textit{1}\textit{0.3 nm.} This corresponds to effective sub-dimer formations of the initial AuNPs \textit{(80 nm)}.\\
A solvent medium specific scattering background that is observed for the \textit{0.17 mM} DMAE modified sample (Cytometry, see Fig. \ref{fig:abb3}) and for the agglomerated control sample in the flow-cytometry measurements vanishes under plasmonic enhancement.\\

This is due to the homogenization according to a near resonant effective size of the AuNPs formation in the stabilized \textit{1.7 mM DMAE} AuNP solution. The nonspecific scattering due to cytometry measurement background is successfully suppressed using the novel approach of volume fluorescent illumination flow-cytometry diagnostics presented in this article.\\
This suppression of "noise" is seen to be driven by a change of the spatial characteristic of the fluorescent emission. This change is based on a direct correlation of the strongly localized scattering process in presence of a fluorescent medium. The fluorescence of the solvent medium is getting directed sideward during the propagation in the scattering volume of the samples. This process is found to be dependent on the diffuse character of the illumination surrounding the particles on a molecular scale. A specificity of detection to the sample response is guaranteed by the spectral separation of the \textit{FL} emission in our cytometry measurements. The measurements demonstrate a proof of principle for this new technique.

\subsection{Simulation of Volume Scattering Light Transport}

For a deeper analysis of the underlying processes simulations of the transport mechanisms in the scattering sample volume are performed to gain information on the spatial shift of the sideward scattering. The proposed volume like behavior of the scattering provides the key-mechanism to result in a sideward direction from the cytometry cell that is specific for the scattering particles.\\
Thus, the dependence of the sideward component of the AuNP scattering in the solvent volume is studied in our model. We compare the volume transport of the diffuse illumination and the transport of individual scattering events specific for particle scattering. The volume scattering is supposed to be separated from the plasmonic resonance. The light transport calculations are performed for an (off-resonant) effective size of \textit{1.83 $\mu$m} of the AuNP compounds, based on our analysis of the \textit{0.17 mM} AuNP solution. The corresponding optical scattering in a model system is implemented by using an analytical \textit{Volume Scattering Function (VSF)} based on a \textit{Henyey-Greenstein (HG)} scattering phase and an anisotropy parameter of \textit{g=0.64}. This represents the forward- to backward anisotropy of the scattering events and is retrieved from MIE calculations. The scattering coefficient is set to a moderate value of $\mu_{s}$ \textit{=139.861 1/cm}. This corresponds to a concentration of \textit{3} primary AuNPs \textit{(80 nm per cubic-micron)}. The volume fluorescence is included as a transport carried by isotropic \textit{(g=0)} scattering events. The "scattering" coefficient, which is associated to these fluorescence events, is assumed to be equal to the particle scattering $\mu_{s}$.\\

\begin{figure}

\includegraphics*[width=\columnwidth]{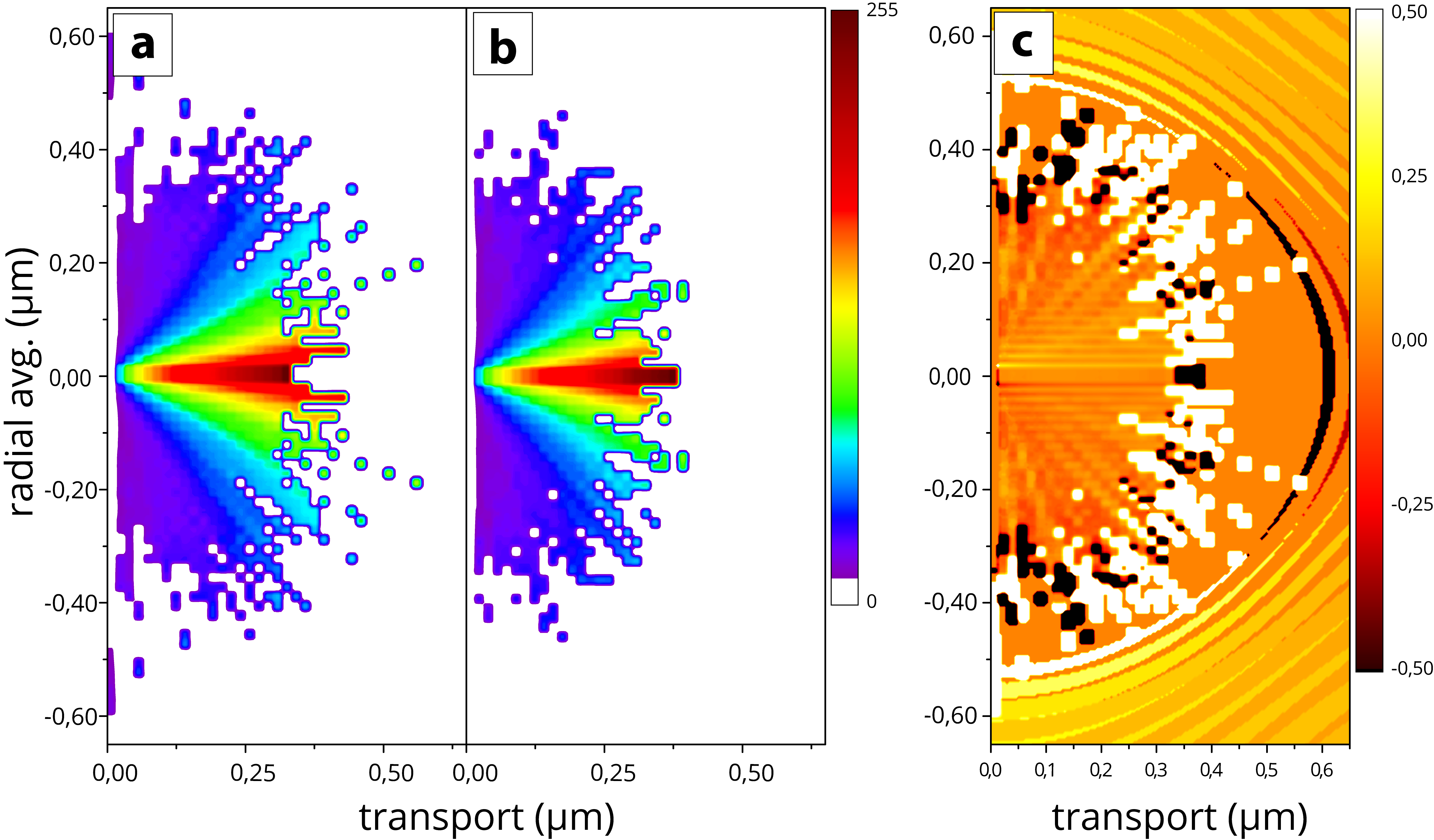} 

  \caption{\footnotesize{Volume Light Transport simulated maps of the attenuation for g=0 (a) and g=0.64 (b) due to the scattering events, where Henyey-Greenstein (HG) analytic phase functions are used forming the Volume Scattering Functions (fluorescence, isotrope $g_0$ and effective particle scattering $g_{64}$). The scattering coefficient and the anisotropy parameter $g$ are used as they are retrieved from MIE calculations. A density of 3 primary AuNPs per cubic micron and the highest observed agglomerate sizes of about 1830 nm are imposed. (c) asymmetry analysis of (a) and (b) following the equation $\frac{g_0 - g_{64}}{g_0+g_{64}}$.  
}}
  \label{fig:abb4}
\end{figure}

In Fig. \ref{fig:abb4} a, b the spatial characteristics of the volume fluorescence \textit{(g=0)} and the fundamental scattering \textit{(g=0.64)} are shown. The volume cells are parameterized by giving a radial axis and a sideward view dimension to obtain a valid mapping between theory and experiment. The radial direction is oriented perpendicular to the incident illumination. The transport dimension in sideward direction corresponds to the \textit{SSC} and \textit{FL} views of the detection in our flow-cytometry measurements. A "measurement" correlated to a sideward collimation in the cytometry geometry is done by averaging both maps retrieved from simulation along the radial cell dimension, \textit{(length 1.3 $\mu$m)}, respectively.\\


Fig. \ref{fig:abb4} c shows an asymmetry analysis of the 8-bit representation of the volume light transport simulation. The data from Fig. \ref{fig:abb4} a and b are processed according to the following equation:
\begin{equation}
\frac{g_0 - g_{64}}{g_0+g_{64}}
\end{equation}
where $g_0$ represents the matrix from Fig. \ref{fig:abb4} a and $g_{64}$ the matrix from Fig. \ref{fig:abb4} b.\\

The given colorscale supports a direct representation of the dominating volume transport direction for $g_0$ (orange to white) and $g_{64}$ (orange to black). One has to keep in mind that the definition of asymmetry used here removes any kind of absolute amplitude values explaining the ring like structures at long radial distance from zero position. Fig. \ref{fig:abb4}c clearly shows the stronger attenuation from the agglomerated sample ($g_{64}$) comparing the amplitudes from Fig. \ref{fig:abb4} a and b with the dominating occurrence of black areas with respect to the radial dimension. Up to a transport depth of 0,2 $\mu$m the $g_{64}$ contributions are significantly stronger taking the corresponding amplitude values of (b) into account. Thus the dimension of the spread along the axis of illumination is found to depend on the pure volume fluorescence ($g_0$, white) or the particle volume scattering ($g_{64}$, black) in the cell.\\


Due to volume scattering events the attenuation of the transport depth is reduced with respect to the pure volume fluorescence. As a consequence one can clearly see that a significant decrease of the optical divergence of the sideward emission from the sample volume. From these simulations we can demonstrate that a well defined particle dependent specificity exists even under the assumption of fluorescent illumination in the volume. Experimentally this specificity is supported by a well defined cone of acceptance given by the objective lens in sideward detection direction in the FCM. Furthermore the widely spread emission of the fluorescent illumination in the cytometry measurements is associated to this cone of collimation only when scattering by the AuNPs is taken into account.


As a consequnce and supported by the simulations a higher degree of accuracy of the FCM diagnostics in our novel approach is proposed to be achieved if the numerical aperture of the sideward detecting collimating objective lens is reduced by an aperture-stop. The radial dimension of the model cell of \textit{1.3 $\mu$m} corresponds to a \textit{NA=0.53} for the collimation \textit{(x50, WD 0.4 mm)}. To optimize the specification of a novel FCM5 diagnostic under ambient illumination to a broad range of particle sizes this represents an initial value to match the collimation aperture as a key-parameter modifing a FCM device. Thus further measurements are planned to verify this proposal in more detail.

\section{Conclusions}

In this article we demonstrate a significant improvement of cytogram feature separation quality by a novel flow-cytometry diagnostic approach. A diffuse illumination of the solutions was applied and the analysis revealed the volume scattering processes to drive the specificity of the particle scattering events.\\
The enhancement of the cytogram quality and feature separation is successfully demonstrated by comparison of fluorescence \textit{FL} to fundamental scattering \textit{SSC} measurements. The volume fluorescence of the solvent medium provides the illumination in these measurements. Thus no apparative changes of the utilized conventional FCM device is needed and the measurement can be performed by the spectral separation of fluorescence in the sideward detection pathway of the device.\\
For initial measurements citrate-capped 80 nm sized gold nanoparticles were stabilized in aqueous solution by chemical reduction providing 2-(dimethylamino)ethanol tertiary amine functionality. Detailed studies of the dispersities of the particle sample solutions are performed. Even under presence of an unspecific scattering background resulting from the solution, we can demonstrate a significant enhancement in data quality opening a broad range of new applications.\\
The nature of the volume scattering mechanism is studied in more detail based on Monte-Carlo modeled spread functions. A reduced divergence in the sideward emission from the cytometry cell is verified to rely on the specific scattering events of the particle in solution. The model provides information to associate this to the cone of optical acceptance of the device in sideward detection.\\
A variation of the aperture of the FCM detection in this direction is identified as another key-parameter to further optimization of the performance and accuracy in the scattering compound separation by the flow-cytometry approach discussed here.

\section{Acknowledgements}

We gratefully acknowledge Juliane Zirm (Institute for Bioprocessing and Analytical Measurement Techniques, Heilbad Heiligenstadt, Germany) for technical assistance. We are indebted to Matthias Steinberg (Sysmex Partec GmbH, PMU, M\"unster, Germany) for fruitful discussions and providing detailed technical specifications of the \textit{CyFlow Space} device as well as to Lilli Schneider (Electronic Structure, Department of Physics, Osnabr\"uck University, Germany) for microbiological scientific knowledge supports.\\

\bibliography{bib} 


\end{document}